\documentclass[aps,twocolumn,showpacs,floats,superscriptaddress]{revtex4}
\usepackage{graphicx}
\usepackage{amssymb,amsmath}
\usepackage{color}
\usepackage{bm}
\begin{document}

%%%%%%%%%%%%%%%%%%%%%%%%%%%%%%%%%%%%%% AUTHORS %%%%%%%%%%%%%%%%%%%%%%%%%
\author{Nikolay V. Prokof'ev}
\affiliation{Department of Physics, University of Massachusetts,
Amherst, MA 01003, USA}
\affiliation{National Research Center ``Kurchatov Institute,''
123182 Moscow, Russia}

\author{Boris V. Svistunov}
\affiliation{Department of Physics, University of Massachusetts,
Amherst, MA 01003, USA}
\affiliation{National Research Center ``Kurchatov Institute,''
123182 Moscow, Russia}
\affiliation{Wilczek Quantum Center, School of Physics and Astronomy and T. D. Lee Institute, Shanghai Jiao Tong University, Shanghai 200240, China}

%%%%%%%%%%%%%%%%%%%%%%%%%%%%%%%%%%%%%%%%%%%%%%%%%%%%%%%%%%%%%%%%%%%%%%%%%%%%%%

\title{Algebraic Time Crystallization in a Two-dimensional Superfluid}

%%%%%%%%%%%%%%%%%%%%%%%%%%%%%%%%%%%%%%%%%%%%%%%%%%%%%%%%%%%%%%%%%%%%%%%%%%%%%%

\date{\today}
\begin{abstract}
Time crystallization is a hallmark of superfluidity, indicative of the fundamental fact that along with breaking the global U(1) symmetry,  superfluids also break time-translation symmetry.
While the standard discussion of the time crystallization phenomenon is based on the notion of the 
global phase and genuine condensate, for the superfluidity to take place in two dimensions an algebraic (topological) order is sufficient. We find that the absence of long-range order in a finite-temperature two-dimensional superfluid translates into in an algebraic time crystallization caused by
the temporal phase correlations. The exponent controlling the algebraic decay is a universal function of the superfluid-stiffness-to-temperature ratio; this exponent can be also seen  in the power-law singularity of the Fourier spectrum of the AC Josephson current.
We elaborate on subtleties involved in defining the phenomenon of time crystallization in both 
classical-filed and all-quantum cases and propose an experimental protocol in which the broken time 
translation symmetry---more precisely, temporal correlations of the relative phase, 
with all possible  finite-size, dimensional, and quantum effects included---can be observed 
without permanently keeping two superfluids  in a contact.

\end{abstract}


% 03.75.Kk  Dynamic properties of condensates; collective and hydrodynamic excitations, superfluid flow

% 03.75.Lm  Tunneling, Josephson effect, Bose-Einstein condensates in periodic potentials, solitons, vortices, and topological excitations (see also 74.50.+r Tunneling phenomena; Josephson effects in superconductivity)

% 74.50.+r   Tunneling phenomena; Josephson effects (for SQUIDs, see 85.25.Dq; for Josephson devices, see 85.25.Cp; for Josephson junction arrays, see 74.81.Fa)

% 11.30.-j	 Symmetry and conservation laws

\maketitle

In the theory of superfluidity, the phenomenon of spontaneously
breaking the time-translation symmetry---{\it time crystallization} (TC),
in modern terminology \cite{Wilczek,Shapere}---is inseparable from the phenomenon of
spontaneously breaking the global U(1) symmetry. Indeed, the very fact of existence
of the global superfluid phase, $\Phi$, implies
that $\Phi$ evolves in time, obeying the universal Beliaev--Josephson--Anderson
relation (in the reference frame of the normal component)
\begin{equation}
\dot{\Phi} = -\mu ,
\label{phi_dot}
\end{equation}
with $\mu$  the chemical potential (in the units of $\hbar$ for bosonic systems) \cite{SBP}.
Any periodic function of $\Phi$ then is also a periodic function of time.
Relation (\ref{phi_dot}) also applies to the hydrodynamics of superfluids with immobile normal component (being considered below),
in which case both $\Phi$ and $\mu$ are understood as smooth time-dependent fields.
Since superfluidity is essentially the classical-field phenomenon \cite{SBP},
here and throughout all our calculations, we use classical hydrodynamic variables.
Quantum mechanics, while bringing about some interesting finite-size effects
(e.g., the  macroscopically slow quantum diffusion of the superfluid phase),
 is fully consistent with the classical-field picture despite
 widespread misconceptions, as we discuss later.

There are many alternative ways to reveal the TC phenomenon. One
prominent manifestation is the alternating current (AC)  Josephson effect, when the matter flux,
$J$, between two weakly coupled superfluids, $A$ and $B$, with different chemical potentials,
$\mu_A$ and $\mu_B$, oscillates in time ($J_0$ is the Josephson coupling constant) \cite{Josephson}:
\begin{equation}
J = J_0 \sin  \Delta \Phi ,
\label{Josephson_a}
\end{equation}
\begin{equation}
\Delta \Phi = \Phi_A(t) - \Phi_B(t)= \Delta \mu \, t , \qquad \Delta \mu= \mu_B - \mu_A.
\label{Josephson_b}
\end{equation}
The corresponding microscopic system parameters such as interactions,
density, or temperature, can be either equal or (very) different; i.e. there is no need
to assume that the two systems are identical and only differ by an energy offset.

Since Wilczek's remark in Ref.~\cite{Wilczek}, it became common to argue \cite{Watanabe}
that the AC Josephson effect is fundamentally nonequilibrium/driven and as such does not
qualify for the phenomenon of TC. To begin with, AC Josephson effect is nothing but an
experimental probe of the phenomenon that exists regardless of the method used to study it;
there is nothing special about the probe being weakly dissipative.
The classical-field picture readily makes it clear: All the quantities in
Eqs.~(\ref{phi_dot}) and (\ref{Josephson_b}) remain perfectly well defined in the absence
of any direct contact between two superfluid systems.
Correspondingly, TC can be also revealed by either releasing
tiny fractions of atoms from each system and recording matter wave interference patterns \cite{Castin}
separated by some amount of time $t$, or by switching on the
Josephson link between the two superfluids only for two infinitesimally short
time intervals separated by some waiting time $t$. During the waiting time both
superfluids---fully equilibrium on their own---evolve absolutely independently.

Nevertheless, even at the classical-field level, we face the problem of
not being able to measure the phase or the chemical potential of the system directly while
keeping the particle number constant. Under these conditions,
the phase-rotation period in an isolated superfluid depends on the choice of the energy offset.

In its conventional form of Eqs.~(\ref{Josephson_a})--(\ref{Josephson_b}), the AC Josephson effect requires the genuine long-range order---the existence of the global phases $\Phi_A(t)$ and $\Phi_B(t)$---whereas  for the superfluidity to take place,
the topological (aka algebraic) order is sufficient \cite{Berezinskii, K_T}. In a superfluid with the algebraic order, the off-diagonal correlations of the order parameter decay algebraically. It is then natural to expect that in such superfluids the phenomenon of {\it algebraic time crystallization} can take place, manifesting itself in
a power law (as opposed to exponential law in a normal state) decay of the temporal  Josephson current oscillations.

In this Letter, we present the corresponding theory for a two-dimensional finite-temperature
disordered superfluid. The presence of disorder renders the
picture especially simple and clean by removing the trivial hydrodynamic mode
associated with translational motion of the normal component.
We show that the exponent controlling the algebraic decay of the temporal Josephson current oscillations is a sum of two individual exponents characterizing the spatial algebraic order in each of the two systems. This exponent---coming from temporal dephasing in each of the two superfluids, cf.~\cite{Burkov}---also provides direct access to the superfluid density, i.e. the quantity that cannot be expressed in terms of local in space observables, but nevertheless, can be measured using purely local in space probes.

{\it Calculations.} The absence of the long-wave translational motion of the normal component in a disordered superfluid immediately implies the absence
of entropy transport, or even a time evolution, as long as the heat transport processes can be neglected. The latter will
be assumed in what follows. The hydrodynamics is then fully described by two fundamental relations, the first one being Eq.~(\ref{phi_dot}).
The second relation is between the superfluid current density, ${\bf j}_s$, and the gradient of superfluid phase:
\begin{equation}
{\bf j}_s = \Lambda_s   \nabla \Phi ,
\label{j_s}
\end{equation}
with  $\Lambda_s$ the superfluid stiffness.
The conservation of matter implies that the superfluid current density
obeys the continuity equation in terms of the time-derivative of the total density $n$:
\begin{equation}
\dot{n} + \nabla \cdot {\bf j}_s =0 \qquad \Rightarrow \qquad \dot{n} +
\nabla \cdot (\Lambda_s   \nabla \Phi  )=0 .
\label{cont}
\end{equation}

In what follows, we confine ourselves to the most simple and natural case of constant---both in space and time---entropy density.
At this fixed value of entropy density, any two out of the four fields, $\nabla \Phi$, $\mu$, $n$, and $\Lambda_s$ are related to the other two
by thermodynamic equations of state. We then select the pair
$(\nabla \Phi, \mu)$ to be the set of prime hydrodynamic fields (the fields
$n\equiv n(\nabla \Phi, \mu)$ and $\Lambda_s\equiv \Lambda_s(\nabla \Phi, \mu)$
are related to them, at the given fixed value of entropy, by the corresponding
equations of state), and linearize around the equilibrium point
$(\nabla \Phi =0, \mu = \mu_0)$. By introducing additional thermodynamic
quantities through partial derivatives of density evaluated at $(0,\mu_0)$
\begin{equation}
\varkappa = {\partial n \over \partial \mu},
\quad \qquad {\bf Q} = {\partial n \over \partial \nabla \Phi} =0,
\label{kappa_and_Q}
\end{equation}
(the last equality follows from system's isotropy),
we re-write the linearized continuity equation as
\begin{equation}
\varkappa \dot{\mu}\,  +\,  \Lambda_s\,  \Delta \Phi = 0.
\label{cont2}
\end{equation}
Here and in what follows, $\Lambda_s\equiv \Lambda_s(0,\mu)$.

For the sound modes---describing the long-range fluctuations of the field
$\Phi$---we introduce the (small) fields $\mu_1$ and $\Phi_1$, such that
\begin{equation}
\mu = \mu_0 + \mu_1 , \quad \qquad \Phi = -\mu_0 t + \Phi_1 \, .
\label{mu_1_and_Phi_1}
\end{equation}
From (\ref{cont2}) we have
\begin{equation}
\dot{\mu}_1  + c^2 \Delta \Phi_1 = 0, \quad \qquad c^2 = {\Lambda_s \over \varkappa } \;.
\label{sound_velocity}
\end{equation}
Since (\ref{phi_dot}) and (\ref{mu_1_and_Phi_1}) imply $\dot{\Phi}_1 = -\mu_1$,
we conclude that both $\mu_1$ and  $\Phi_1$ obey the wave equation with the sound velocity $c$.

For the long-wave dynamics of the field $\Phi$ we thus get
\begin{equation}
\Phi({\bf r},t) = -\mu_0 t + \sum_{\bf k}^{k<k_*} \,\Phi_{\bf k} \, e^{i({\bf k}\cdot {\bf r} -  ck t)} \;,
\label{Phi_long_wave}
\end{equation}
where $k_*$ is a certain cutoff.
At the thermodynamic equilibrium, the averages of the squares of amplitudes $\Phi_{\bf k}$
are given by
\begin{equation}
\langle \, | \Phi_{\bf k} |^2 \rangle = {T\over \Lambda_s k^2} .
\label{Gibbs}
\end{equation}
Relations (\ref{Phi_long_wave}) and (\ref{Gibbs}) is all we need to characterize the dephasing of the Josephson current.

Our starting point is the standard Josephson relation (\ref{Josephson_a}) with two important modifications
specific for the superfluid state in two dimensions. First, the phase difference across
the weak link should be now expressed in terms of fluctuating {\it fields}
evaluated at the weak link position, $r=0$:
$\Delta \Phi (t) = \Phi_A(r=0,t) -\Phi_B(r=0,t)$.
This quantity explicitly depends on the cut-off $k_*$, see Eq.~(\ref{Phi_long_wave}).
Second, the Josephson coupling $J_0$ also becomes a cut-off dependent quantity. Obviously,
physical answers should not depend on an arbitrarily chosen large wavelength scale.

We rewrite Eq.~(\ref{Josephson_a}) as [with $J_0\equiv J_0(k_*)$]
\begin{equation}
J(t) = J_0 \sin [\Delta \Phi(t)] = J_0  \sin [\Delta \mu t + \Delta \Phi_1(t)]  \, ,
\label{Josephson2}
\end{equation}
where, by construction, we only keep the slow dynamics of the phase field, and assume that
dynamic fluctuations happening at time scales shorter than $t_* = 1/ck_*$ cannot be resolved
and thus are absorbed in the definition of $J_0$.  Given that $J(t)$ is time dependent
(on top of expected oscillations) we compute the current-current correlation function
\begin{equation}
C(t)= \langle \langle J(t_0) J(t_0+t) \rangle_{t_0} \rangle = \langle  J(0) J(t) \rangle \;,
\label{JJ1}
\end{equation}
where the last average is taken over the equilibrium state of the system (the average over the
``initial" time $t_0$ is dropped because equilibrium correlation functions may only depend
on the time difference). Substituting here Eq.~(\ref{Phi_long_wave}) and performing Gaussian
averages of linear exponentials [with mean square fluctuations given by (\ref{Gibbs})], we obtain
\begin{equation}
C(t)= J_0(t_*)^2 \cos (\Delta \mu \, t ) \, \left( \frac{t_*}{t} \right)^{\alpha }  \,,
\label{JJ2}
\end{equation}
\begin{equation}
\alpha=\frac{1}{2\pi} \left( \frac{T_A}{\Lambda_s^{(A)}} + \frac{T_B}{\Lambda_s^{(B)}}  \right) .
\label{alpha}
\end{equation}
In this derivation, one should pay attention to the following circumstances: (i) we have two independent superfluid fields, (ii)
harmonics ${\bf k}$ and $-{\bf k}$ are not independent. The summation over harmonics is performed
with logarithmic accuracy. We also consider the possibility that superfluids A and B can be
at different temperatures and have different microscopic properties. Recalling that at the Berezinskii--Kosterlitz--Thouless transition,
the Nelson--Kosterlitz relation, $T/\Lambda_s = \pi/2$, takes place, we see that the value of the exponent $\alpha $ can be as large as $1/2$.

If we were to increase the cutoff time $t_* \to t_*'$ and absorb additional short-time fluctuations
into the definition of $J_0$, we would obtain the same expression with
$J_0(t_*) \to J_0(t_*')=J_0(t_*)(t_*/t_*')^{\alpha/2}$. The power law dependence of the Josephson
current amplitude on the cutoff scale, $J_0 \propto 1/t_*^{\alpha/2}$, implies that in the static
thermodynamic limit this amplitude vanishes as $J_0 \propto 1/L^{\alpha/2}$, where $L$ is the linear
system size. In other words, in the static limit the Josephson effect in two-dimensional systems
is absent \cite{Hermele}.

{\it Measuring the time crystallization phenomenon}.
There are two aspects regarding measurements of the superfluid phase and TC.
The first one is purely quantum, while the second one is purely classical;
both are, in fact, generic to any case of spontaneous symmetry breaking.
For an isolated quantum superfluid with a given number of particles, the uncertainty
principle and U(1) symmetry dictate that we are dealing with the equal-weight superposition of fixed-phase
states. Hence, to create a state with well-defined phase one needs to perform a projective
measurement of the appropriate physical property, such as the Josephson current.
Qualitatively, this situation is in a one-to-one correspondence with the case of an isolated quantum
crystal with fixed momentum in a box with periodic boundary conditions. By the uncertainty principle
and translational symmetry, the expectation value for the center-of-mass coordinate is a uniform function,
as in a liquid. A measurement is required to produce a state that one normally has in mind when thinking of a crystal.

The second, purely classical aspect, comes from the fact that whenever we are dealing
with spontaneous breaking of a continuous symmetry respected by {\it all} systems in the Universe,
the value of an observable quantifying the symmetry breaking can be measured only {\it relative}
to the corresponding value of some reference system. For example, if we were to mention the position
${\bf r}_0$ of an atom in a classical crystal, we would only be able to do this with respect to a
certain {\it material} reference frame; in other words, ${\bf r}_0$ represents a {\it relative}
rather than an {\it absolute} coordinate.  The classical aspect of measuring the superfluid phase
is precisely like that: only {\it relative} values of the superfluid phase can be meaningfully measured.
The only extra restriction is that both the system and its reference must contain superfluid states
composed of the same type of matter that can flow between the two.

Recently, Watanabe and Oshikawa presented a ``no go" theorem  \cite{Watanabe}
establishing the absence of TC in the ground and finite-temperature states
of macroscopic systems for which energy is the only thermodynamically relevant constant of motion.
Neither classical nor quantum superfluids satisfy the conditions of the theorem because
of the total amount of matter conservation law. The freedom of adding to
the system's Hamiltonian any function of conserved quantities formally allows one
to suppress the breaking of the time-translation symmetry in a particular isolated superfluid,
but this cannot be achieved {\it for all superfluids simultaneously}!

The authors of  Ref.~\cite{Watanabe} also argued that the absence of TC in a quantum superfluid
is due to the fact that the phase is fundamentally ill-defined when the number of particles is fixed.
All by itself this is not a problem because all it takes to make the phase well defined is to measure it
relative to the reference system! Moreover, revealing temporal phase correlations does not
even require measuring the phase. The situation here is precisely analogous to measuring the
single-particle density matrix: there is no need for the state of the system to have a nonzero
expectation value for the field operator. In the next section, we present an explicit
experimental protocol revealing temporal phase correlations in the equilibrium quantum
state without pre-collapsing it onto a state with a well defined phase.
The protocol allows one to experimentally demonstrate that a pair of equilibrium superfluids
demonstrates time crystallization.

{\it  Experimental protocol.}  Cold atomic systems apparently provide the best platform
for measuring the algebraic TC effect.
Conceptually, all one needs is to establish a weak link between two superfluid states.
Creating multiple well-isolated two-dimensional superfluids was already achieved for
phase interference experiments some time ago \cite{Zoran,Stock}, while more recent work on superfluid
Fermi gases \cite{Thierry} has established the possibility of measuring transport through point contacts.
As a natural part of the algebraic TC effect, one establishes a way to measure superfluid density
using purely local in space probe.

Here we propose an experimental protocol of measuring temporal phase correlations between two disconnected finite-size quantum superfluids that fully preserves the quantum uncertainty of the relative phase
between the two samples:

$\bullet$ Prepare two isolated superfluid samples at close but different chemical potentials.

$\bullet$  At $t=0$, connect the two samples by a local weak link for a short,
           $\Delta t \, \Delta \mu \, \ll \, 1$, period of time.

$\bullet$  Keep samples isolated for a much longer time interval $t \, \Delta \mu \, > \, 1$ and
           then repeat the previous step.

$\bullet$ Quickly, on time scales $\, \ll \, 1/\Delta \mu$, apply a deep optical lattice to localize all
          atoms in the system and count atom numbers $N_A$ and $N_B$ using single-site microscopy \cite{Bloch_2010,Greiner_2010}.

Repeating the protocol many times under identical conditions allows one to accumulate
representative statistics and process the data with the help of an auxiliary experimental run
that skips the next-to-last step of the above-described protocol.
The outcome of the auxiliary run is the expectation value $\bar{N}_{AB} = \langle N_A - N_B \rangle$
that averages typical particle number differences taking place {\it before} the two samples are
disconnected for a period of time $t$.  The key statistical observable is then
\begin{equation}
K(t) \, =\,  \langle \; (N_A(t) - N_B(t)  - \bar{N}_{AB})^2  \, \rangle .
\label{K}
\end{equation}
This quantity is the sum of two fundamentally different dispersions of the random number $N_A(t) - N_B(t)  - \bar{N}_{AB}$: one is characterizing irreproducibility of the initial state preparation,
and the other is reflecting temporal current-current correlations, the all-quantum counterpart of (\ref{JJ1}).
The latter dispersion reveals the time crystallization effect, while the former dispersion is $t$-independent and thus creates no problem except for that of a signal-to-noise ratio, which can be improved by collecting more statistics and optimizing setup parameters. To ensure that the time-dependent contribution to
dispersion is large, one needs to have $J_0(t_*)/\Delta \mu  \gg 1$.

A reader familiar with the method of measuring temporal phase correlations proposed earlier by Kurkjian, Castin, and Sinatra (KCS) may wonder whether our protocol
has any advantage compared to the KCS one, especially given the fact that the latter deals with a singe sample, while ours needs two. Here we observe that working with two samples
significantly enhances the (otherwise quite small) exponent of the algebraic dephasing. A more fundamental question is whether the KCS protocol---as opposed to ours---yields the
 {\it absolute} value of the phase rotation rate. The answer is ``no,'' because the rate is measured with respect to the energy of a single atom at rest. Setting this energy to zero is merely
 a (reasonable) convention.

We acknowledge instructive discussions with Krzysztof Sacha, Rosario Fazio,  and Yvan Castin.
This work was supported by the National Science Foundation under the grant DMR-1720465
and the MURI Program ``New Quantum Phases of Matter'' from AFOSR.
The authors are grateful to the Mainz Institute for Theoretical Physics for its hospitality and partial support during the
completion of this work.

\end{document}